\documentstyle[preprint,aps]{revtex}

\newcommand{\be}{\begin{eqnarray}}
\newcommand{\e}{\end{eqnarray}}

\newcommand{\ov}{\over}
\newcommand{\k}{\kappa}
\newcommand{\p}{\perp}
\newcommand{\s}{\sigma}

\begin{document}
\draft
\tighten
\title{ Twist Four Longitudinal Structure Function for a Positronium-like
Bound State in Light-Front QED}
\author{{\bf Asmita Mukherjee}\thanks{e-mail: asmita@theory.saha.ernet.in}\\ 
Saha Institute of Nuclear Physics, 1/AF, Bidhan Nagar, Calcutta 700064 India}
\date{July 10, 2001}
\maketitle
\begin{abstract}
To have an analytic understanding of the higher-twist structure functions, 
we calculate twist four
longitudinal structure function for a positronium-like bound state in weak
coupling light-front QED. We find that in the weakly
coupled system, the fermionic part of $F_L$ is related to
the kinetic energy of the fermions and not to the interaction. We verify a
previously proposed sum rule in this limit, which 
in this case reduces to a relation connecting the kinetic and the potential
energies to the binding energy of positronium.
Using the analytic form of the wave function of positronium in this limit,
we show that the constituent counting rule does not hold for $x \rightarrow
1$. The twist four $F_L$ in this limit is similar in form to a widely used
phenomenological ansatz.
\end{abstract}
\vskip .2in
{\bf 1. Introduction}
\vskip .2in
Higher twist or power suppressed contributions to deep inelastic scattering
structure functions involve non-trivial non-perturbative information about
the structure of hadrons. Recent experimental results indicate that these 
higher twist effects play an important role in the kinematical range of the
SLAC experiments and it is important to have a clear physical picture of
them. Light-front Hamiltonian QCD offers a theoretical tool to investigate
the deep inelastic scattering structure functions. This is based on physical
intuitions and at the same time employs well defined field theoretical
calculational techniques. The structure functions are expressed as the
Fourier transform of the matrix elements of light-front bilocal currents. 
Fock space expansion of the target state allows us to express these in terms of
light-front multiparton wave functions. 
The interesting aspect of this formulation is that
both perturbative and non-perturbative issues can be addressed within the
same framework \cite{hari1}. Recently, the twist four part of the longitudinal structure
function $F_L$ and the transverse polarized
structure function $g_T$, which is a twist three contribution have been
analyzed in this approach and various interesting issues associated with
them have been addressed
\cite{fl,gt}. The structure functions can be calculated once the light-front
bound state wave functions are known.  
However, the actual calculation for a QCD bound
state in $3+1$ dimension is highly complicated and it requires the recently
developed similarity renormalization techniques \cite{sim} for the light-front QCD
Hamiltonian. The spectra of the heavy quark bound states like charmonium and
bottomonium have been investigated using this technique \cite{mar}. 
The analytic form of the wave function  is not obtained so far.  
This is because, the similarity renormalization
technique generates an additional confining interaction in the effective
Hamiltonian  which makes it impossible to solve the effective
Hamiltonian analytically even in the leading order in bound state Hamiltonian 
perturbation theory.
In this work, we have performed a much simpler but quite interesting analysis.
We have calculated the twist-four part of the longitudinal structure
function for a positronium-like bound state in light-front
QED in the weak coupling limit. In fact, in this limit, QCD results are not
expected to differ too much from the QED results. As a result, this analysis 
is important since it tests
and illustrates the approach in QCD. The advantage here is that, in the
leading order in bound state perturbation theory, the bound state equation
can be solved analytically and the wave function is known. This allows an 
analytic understanding of the problem.                  

In the weak coupling limit, we show that our previously proposed sum rule
\cite{fl} 
reduces to a relation connecting the kinetic and potential energies to the binding energy of
positronium. Using the analytic wave
function of positronium in this limit, we show that the structure functions
fall faster than that predicted by constituent counting rule near $x
\rightarrow 1$. Also, we find that the twist four $F_L$ in the weak coupling
limit has a form 
which is similar to a widely used  phenomenological formula, which connects
the twist four distribution to the twist two distribution.

\vskip .2in
{\bf 2. Twist Four Longitudinal Structure Function}
\vskip .2in
We consider a positronium-like bound state $\mid P \rangle$ given by
\be
\mid P \rangle = && \sum_{\sigma_1, \sigma_2} 
\int {dk_1^+ d^2 k_1^\perp \over \sqrt{2 (2 \pi)^3 k_1^+}} 
\int {dk_2^+ d^2 k_2^\perp \over \sqrt{2 (2 \pi)^3 k_2^+}} 
\nonumber \\
&& \phi_2(P \mid k_1, \sigma_1; k_2, \sigma_2) \sqrt{2 ((2 \pi)^3 P^+}
\delta^3(P-k_1-k_2) b^\dagger(k_1, \sigma_1) d^\dagger(k_2,\sigma_2) \mid 0
\rangle \nonumber \\
&&~~~~~~ + \sum_{\sigma_1,\sigma_2,\lambda_3} 
\int {dk_1^+ d^2 k_1^\perp \over \sqrt{2 (2 \pi)^3 k_1^+}} 
\int {dk_2^+ d^2 k_2^\perp \over \sqrt{2 (2 \pi)^3 k_2^+}} 
\int {dk_3^+ d^2 k_3^\perp \over \sqrt{2 (2 \pi)^3 k_3^+}} 
\nonumber \\
&&~~~~~~~~ \phi_3(P \mid k_1, \sigma_1; k_2, \sigma_2; k_3, \lambda_3)
\sqrt{2 (2 \pi)^3 P^+} \delta^3(P-k_1 -k_2 -k_3) 
\nonumber \\
&&~~~~~~~~~~~~~~~~~~b^\dagger(k_1 ,\sigma_1)
d^\dagger (k_2, \sigma_2) a^\dagger(k_3, \lambda_3) \mid 0 \rangle. 
\label{e8}
\e

Here $\phi_2$ is the probability amplitude to find an electron and positron in
the positronium, $\phi_3 $ is the probability amplitude to find an electron,
positron and a photon in the positronium. We consider upto three particle
sector.
We introduce Jacobi momenta $(x, \kappa^\perp)$ and the boost invariant
amplitudes, $\psi_2$ and $\psi_3$ \cite{rajen}.

We calculate $F^{\tau=4}_4$ in light-front QED in light front gauge, $A^+=0$ 
using weak coupling
approximation. This includes the
effect of dynamical photon. 
 It can be shown that,
for a weak coupling theory the results are the same as obtained using a non-relativistic
approximation. However, the entire calculation is fully relativistic and
exact in the leading order in bound state perturbation theory \cite{bur}. 

The twist-4 part of the fermionic component of the longitudinal structure
function is given by,
\be
F^{\tau=4}_{L(f)}(x) = {M}_1 + {M}_2,
\e
\begin{eqnarray}
{M}_1~=~{ 1 \over Q^2}~ {x^2 (P^+)^2 \over 2 \pi}~ \int dy^- ~
e^{-{ i \over
2}P^+y^-x}~\langle P \mid \overline{\psi}(y^-) \gamma^- \psi(0)-
\overline{\psi}(0) \gamma^- \psi(y^-) \mid P \rangle,
\label{e10}
\end{eqnarray}
\begin{eqnarray}
{M}_2=-{(P^\perp)^2  \over (P^+)^2}
{ 1 \over Q^2} {x^2 (P^+)^2 \over 2 \pi}~ 
\int dy^- ~e^{-{ i \over
2}P^+y^-x}~\langle P \mid \overline{\psi}(y^-) \gamma^+ \psi(0)-
 \overline{\psi}(0) \gamma^+ \psi(y^-)\mid P \rangle.
\label{e11}
\end{eqnarray}

We shall take the mass of the state $\mid P \rangle$ to be M and the electron
and positron mass to be $m$. In the weak coupling (non-relativistic) limit the helicity
dependence of the wave function factorizes away, so it is sufficient to
consider one helicity sector. Here we shall take the two particle state with
helicities $\s_1$ and $\s_2$ up.

For a positronium state $\mid P \rangle$ given by Eq. (\ref{e8}) we obtain,
\be
{F^{\tau=4}_{L(f)}(x)}_{diag} =&&(M_1)_{diag} + (M_2)_{diag}
\nonumber\\&&~~~~= {4\ov Q^2} \int dx_1 d^2\k_1^\p ((\k_1^\p)^2+m^2) {\mid \psi_2
 \mid }^2 \Big [ \delta (x-x_1) + \delta(1-x-x_1) \Big ] \nonumber\\&&~~~~
+{4\ov Q^2}\sum \int dx_1 d^2
\k_1^\p \int dx_2 d^2\k_2^\p {\mid \psi_3 \mid }^2 \Big [
((\k_1^\p)^2+m^2)\delta(x-x_1) + \nonumber\\&&~~~~~~~~~~~~~~~~~~~~
((\k_2^\p)^2+m^2) \delta(x-x_2) \Big ].
\label{f1}
\e
The off-diagonal contributions to $ F^{\tau=4}_{L(f)}(x)$ comes from $ M_1 $
alone.
\be
(M_1)_{off-diag} = -{4\ov Q^2} {e^2\ov {2(2\pi)^3}}\int dx_1d^2\k_1^\p \int
dyd^2\k^\p \Big [ M_1^a + M_1^b + M_1^c + M_1^d \Big ]
\e
where
\be
M_1^a = {1\ov {(x_1-y)^2E}}\Big [ {2(\k_1^\p)^2 y \ov x_1}\delta(x-x_1) -
{2(\k^\p)^2 x_1\ov y} \delta (x-y) \Big ]
{\mid \psi_2^{\s_1\s_2} (x_1, \k_1^\p)\mid }^2,   
\label{u1}
\e
\be
M_1^b = {1\ov {E(x-y)^2}}\Big [ -(\k_1^\p)^2\delta(x-x_1) + (\k^\p)^2
\delta(x-y) \Big ]  \nonumber\\ ~~~~~~~~~~~~~\Big (
 \psi_2^{*\s_1\s_2} (x_1, \k_1^\p) \psi_2^{\s_1\s_2} ( y,\k^\p) + h.c. \Big
),
\label{u2}
\e
\be
M_1^c = {1\ov {(y-x_1)^2 E'}} \Big [ {2(\k_1^\p)^2y\ov
(1-x_1)}\delta(1-x-x_1) - {2(\k^\p)^2 (1-x_1)\ov y} \delta (x-1+y) \Big ]
\nonumber\\~~~~~~~~~~~~~~~~~~~~~~~~~~~~~~~~~~~~{\mid \psi_2^{\s_1\s_2} 
(x_1, \k_1^\p)\mid}^2,
\label{u3}
\e  
\be
M_1^d = {1\ov {E'(y-x_1)}}\Big [ (\k^\p)^2 \delta (x+y-1) - (\k_1^\p)^2
\delta (1-x-x_1) \Big ]\nonumber\\~~~~~~~~~~~~\Big ( \psi_2^{*\s_1\s_2} (x_1, \k_1^\p)
\psi_2^{\s_1\s_2}( y,\k^\p) + h. c. \Big ).
\label{u4}
\e
In this calculation we have taken all operators to be normal ordered.
Also, in the Eq. (\ref{u1})- (\ref{u4})  we have neglected
all mass terms in the vertex, since these terms are
suppressed in the non-relativistic limit. The energy denominators are given
by,
\be
E=M^2-{{(\k^\perp)^2+m^2}\over y}-{{(\kappa_1^\perp)^2+m^2}\over
1-x_1}-{(\kappa_1^\perp-\k^\perp)^2\over x_1-y}\nonumber\\
E'=M^2-{{(\kappa_1^\perp)^2+m^2}\over x_1}-{{(\k^\perp)^2+m^2}\over
y}-{(\kappa_1^\perp-\k^\perp)^2\over (y-x_1)}.
\label{u5}
\e 

We define the twist four longitudinal photon structure function as \cite{fl},
\be
F_{L(g)}^{\tau=4}(x) && = { 1 \over Q^2} {x P^+ \over 2 \pi} \int dy^- ~
e^{-{i \over 2} P^+ y^- x} \nonumber \\
&& ~~~~~\Big [ \langle P \mid (-) F^{+ \lambda}(y^-) F^-_{\lambda}(0) + 
{ 1 \over 4} g^{+-} F^{\lambda \sigma} (y^-) F_{\lambda \sigma}(0) \mid
P \rangle \nonumber \\
&& ~~~~~~~ - {(P^\perp)^2 \over (P^+)^2} \langle P \mid F^{+ \lambda}(y^-) 
F^+_{\lambda}(0) \mid P \rangle + ( y^- - 0)\Big ].
\e

$ F_{L(g)}^{\tau=4}(x)$ has both diagonal and off-diagonal parts.
We take all operators to be normal ordered and we get, 
\be
{{ F_{L(g)}^{\tau=4}(x)}\ov x}_{diag}&&= {4\ov Q^2}  \int dx_1d^2\k_1^\p \int dy
d^2\k^\p
{\mid \psi_3 \mid }^2 {(-\k_1^\p-\k^\p)^2 \ov (1-x_1-y)} \delta ( 1-x_1-x-y) 
\nonumber\\&&-{4\ov Q^2} {4e^2\ov
{2(2\pi)^3}}\int dx_1d^2\k_1^\p \int dyd^2\k^\p \psi_2^*(x_1,
\k_1^\p)\nonumber\\&&~~~~~~~~~~~~~~~~~~~~~~~~~~~~~~~~~~
 \psi_2(y,\k^\p){1\ov {(x_1 - y)^2}}\delta(x_1-x-y).
\e
The second term in the right hand side is the contribution of the instantaneous
interaction.
The off-diagonal contribution is,
\be
{ F_{L(g)}^{\tau=4}(x)}_{off-diag} = G_1 + G_2
\e
where,
\be
G_1 &&= -{4\ov Q^2} {e^2\ov {2(2\pi)^3}}  \int dx_1 d^2\k_1^\p \int dy
d^2\k^\p  {x\ov E(x_1-y)^2} \nonumber\\&&~\Big [ \Big ( -4{ (\k_1^\p - \k^\p)^2\ov
(x_1-y)} + {2(\k_1^\p)^2\ov x_1} - {2(\k^\p)^2\ov y} \Big ) 
{\mid \psi_2^{\s_1\s_2} (x_1, \k_1^\p)\mid
}^2\nonumber\\&&~~~~~~~~~
+\Big ( {2(\k_1^\p-\k^\p)^2\ov {x_1-y)}} + {(\k_1^\p)^2\ov (1-x_1)} -
{(\k^\p)^2\ov (1-y)} \Big ) 
\nonumber\\&&~~~~~~~~~~~~~~~\Big ( \psi_2^{*\s_1\s_2} (x_1, \k_1^\p)
 \psi_2^{\s_1\s_2} (y,\k^\p) + h.c. \Big )\Big ],   
\delta (x-x_1 +y), 
\e
\be
G_2 &&= -{4\ov Q^2} {e^2\ov {2(2\pi)^3}} \int dx_1 d^2\k_1^\p \int dy
d^2\k^\p  {x\ov {E'(y-x_1)^2}} \nonumber\\&&\Big [\Big (-{4(\k_1^\p -\k^\p)^2\ov
(y-x_1)} - {2(\k^\p)^2\ov (1-y)} + {2(\k_1^\p)^2\ov (1-x_1)} \Big )
{\mid \psi_2^{\s_1\s_2} (x_1, \k_1^\p)\mid
}^2\nonumber\\&&~~~~~~~~~~~~~
+\Big (-{2(\k_1^\p -\k^\p)^2\ov
(y-x_1)} - {(\k^\p)^2\ov y} + {(\k_1^\p)^2\ov x_1} \Big ) 
\nonumber\\ &&~~~~~~~~~~~~~~~~~~~\Big ( \psi_2^{*\s_1\s_2} (x_1, \k_1^\p)
 \psi_2^{\s_1\s_2} ( y,\k^\p) + h.c. \Big )\Big ]
\delta (x_1+x-y).
\e
Where $E$ and $E'$ are given by Eq. (\ref{u5}).

From these expressions, we calculate,
\be
\int_0^1&& {{ F_{L(q)}^{\tau=4}(x) + F_{L(g)}^{\tau=4}(x)}\ov x} dx
= {4\ov Q^2}  \int dxd^2\k^\p {\psi_2}^* \psi_2 [ {(\k^\p)^2\ov
x} + { (\k^\p)^2\ov {1-x}} ]  \nonumber\\&&
+{4\ov Q^2} \sum \int dxd^2\k^\p 
\int dy d^2q^\p {\psi_3}^* \psi_3 [ {(\k^\p)^2\ov
{x}} + { (q^\p)^2 \ov y}+ {(-\k^\p - q^\p)^2\ov (1-x-y)} ]
\nonumber\\&&~~~~~~~+{4\ov Q^2}{e^2\ov {2(2\pi)^3}}\int dxd^2\k^\p 
\int dy d^2q^\p [ A_1 + A_2 + B_1 + B_2] \nonumber\\&&~~~~~~~~
- {4\ov Q^2} { 4e^2\ov {2(2\pi)^3}}\int dxd^2\k^\p \int dy d^2q^\p {1\ov
(x-y)^2} \nonumber\\&&~~~~~~~~~~~~~~~~~~~ \psi_2^*(x,\k^\p)\psi_2( y, q^\p) 
\label{c4}
\e
where,
\be
A_1 = {1\ov E}{1\ov (x-y)}\Big [V_1
{\mid \psi_2^{\s_1 \s_2} (x,\k^\p) \mid }^2 + V_2
\psi_2^{*\s_1 \s_2} (x,\k^\p)\psi_2^{\sigma_1\sigma_2}(y,q^\p)\Big ], 
\label{a1}
\e
\be
A_2 = {1\ov E'}{1\ov (y-x)}\Big [{V'}_2
 \psi_2^{*\sigma_1\sigma_2}
(x,\kappa^\perp)\psi_2^{\sigma_1\sigma_2}(y,q^\p) 
+ {V'}_1
{\mid \psi_2^{\s_1 \s_2} (x,\k^\p) \mid }^2\Big ] 
\label{a2}
\e 
\be
B_1 = {1\ov E}{1\ov (x-y)} \Big [ V_1
{\mid \psi_2^{\s_1 \s_2} (x,\k^\p) \mid }^2 +V_2
\psi_2^{\s_1 \s_2} (x,\k^\p)\psi_2^{*\sigma_1\sigma_2}
(y,q^\p)\Big ] 
\label{c1}
\e
\be
B_2 = {1\ov E'}{1\ov (y-x)} \Big [{V'}_2
\psi_2^{\sigma_1\sigma_2}
(x,\kappa^\perp)\psi_2^{*\sigma_1\sigma_2}
(y,q^\p) + 
{V'}_1{\mid \psi_2^{\s_1 \s_2} (x,\k^\p) \mid }^2\Big ]  
\label{c2}
\e
where $V_1, V_2, {V'}_1$ and ${V'}_2$ are given by,
\be
V_1 = \Big [{2(\k^\p-q^\p)^2\ov (x-y)^2} + {(x+y)\ov (x-y)}\Big ( {(\k^\p)^2 
\ov x^2} +
{{(q^\p)^2}\ov y^2}\Big )\Big ],
\e
\be
V_2 = {V'}_2 = \Big [{(\k^\p)^2(1-2x)\ov {x(1-x)(x-y)}}+
{(q^\p)^2(2y-1) \ov {y(1-y)(x-y)}} -{2(\k^\p-q^\p)^2\ov (x-y)^2} \Big ],
\e
\be
{V'}_1 = \Big [{2(\k^\p-q^\p)^2\ov (x-y)^2} + {(2-x-y)\ov (y-x)
}
\Big ( {(\k^\p)^2 \ov {(1-x)^2}} +
{{(q^\p)^2}\ov {(1-y)^2}}\Big )\Big ].
\e 

 In these
expressions we have kept only those terms in the vertex which survive in the
non-relativistic limit. The helicities $\s_1$ and $\s_2$ are both up and 
 we have neglected all mass terms in the vertex since
in this limit they are suppressed. Also in the weak coupling (non-relativistic) limit, 
we consider only photon exchange interactions and the
off-diagonal terms proportional to ${\mid \psi_2 \mid
}^2 $ originating from self energy effects can be neglected. 

Now, from the expression of $\psi_3$ (see \cite{rajen}) we can write,
\be
\int dxd^2\k^\p \int dy d^2q^\p {\mid \psi_3 \mid }^2 = { e^2\ov
{2(2\pi)^3}} \int dx d^2\k^\p \int dyd^2q^\p [ {1\ov E} B_1 + {1\ov E'} B_2]
\label{c3}
\e
where $B_1$ and $B_2$ are given earlier.
Using this, one can write the second term in the right hand side of Eq. (\ref{c4}) as,
\be
{4\ov Q^2} {e^2\ov {2(2\pi)^3}}&& \int dxd^2\k^\p \int dyd^2q^\p 
[ {1\ov E} B_1 + {1\ov E'} B_2] \nonumber\\&&~~~~~~~~~~
\Big [ {{(\k^\p)^2 + m^2}\ov x} + {{(q^\p)^2
+ m^2} \ov y} + {(-\k^\p-q^\p)^2\ov (1-x-y)} \Big ].
\e
Considering the fact that the total energy is conserved, one can write this
as,
\be
-{4\ov Q^2} {e^2\ov {2(2\pi)^3}}\int dxd^2\k^\p \int dyd^2q^\p
( B_1 + B_2).
\e
So we get,
\be
\int_0^1 {{ F_{L(q)}^{\tau=4}(x) + F_{L(g)}^{\tau=4}(x)}\ov x} dx&&
= {4\ov Q^2} \int dxd^2\k^\p {\psi_2}^* \psi_2 [ {(\k^\p)^2\ov
x} + { (\k^\p)^2\ov {1-x}} ]\nonumber\\&& 
+{4\ov Q^2}{e^2\ov {2(2\pi)^3}}\int dxd^2\k^\p 
\int dy d^2q^\p [ A_1 + A_2 ] \nonumber\\
&&~~~~~~~~~
- {4\ov Q^2} { 4e^2\ov {2(2\pi)^3}}\int dxd^2\k^\p \int dy d^2q^\p {1\ov
(x-y)^2} \nonumber\\&&~~~~~~~~~~~~~
\psi_2^*(x,\k^\p, 1-x, -\k^\p)\psi_2( y, q^\p). 
\e
Also, if we denote ${(\k^\p)^2 + m^2} \ov {x(1-x)}$ by $M^2_o$, then in the
non-relativistic limit, it can be shown that,
$M^2 - M^2_0 \simeq O(e^4)$.
So we neglect this difference in the energy denominators 
and replace the bound state mass in $E$ and $E'$ by
 $ M^2 = {{(\k^\p)^2 + m^2}\over x(1-x)}$. The  energy
denominators then become,
\be
E = {{(\k^\p)^2 + m^2 } \ov x} - {{ (q^\p)^2 + m^2 }\ov y} - {{ (\k^\p -
q^\p)^2} \ov {x-y}} \nonumber\\
= -{1\ov (x-y)} [ ({m\ov x})^2 (x-y)^2 + (\k^\p- q^\p)^2],
\e
\be
E' =  {{(\k^\p)^2 + m^2 } \ov {1-x}} - {{ (q^\p)^2 + m^2 }\ov {1-y}} 
+ {{ (\k^\p -
q^\p)^2 }\ov {x-y}} \nonumber\\
= {1\ov (x-y)} [ ({m\ov {1-x}})^2 (x-y)^2 + (\k^\p-q^\p)^2].
\e
We get, in this limit,
\be
\int_0^1 {{ F_L^{\tau=4}(x)}\ov x} dx&& = {4\ov Q^2} \int dx d^2\k^\p {\mid
\psi_2 \mid }^2 {{(\k^\p)^2 + m^2}\ov {x(1-x)}} \nonumber\\&&
-{4\ov Q^2}{2e^2\ov {2(2\pi)^3}} \int dyd^2q^\p\psi^*_2( x, \k^\p, 1-x,
-\k^\p) \psi_2(y,q^\p,
1-y,-q^\p) \nonumber\\&&~~~~~~~~~~~~~~~~~~~~~~~~~
\Big [ ({m\ov x})^2 {1\ov { (\k^\p-q^\p)^2 + ({m\ov
x})^2(x-y)^2}}\nonumber\\&&~~~~~~~~~~~~~~~~~~~~~~~~~~~~~~~~~~~ + 
      ({m\ov (1-x)})^2 {1\ov { (\k^\p-q^\p)^2 + ({m\ov (1-x)})^2(x-y)^2}}\Big
].
\label{c5}
\e

Here in the weak coupling limit we have omitted the spin indices.

The Fermionic part of the Hamiltonian density is given by,
\begin{eqnarray}
\theta^{+-}_f = i \overline{\psi}\gamma^- \partial^+ \psi 
  = 2 {\psi^{+}}^\dagger \Big [ \alpha^\perp.(i \partial^\perp + gA^\perp) +
\gamma^0m \Big ] {1\over {i \partial^+}} \Big [ \alpha^\perp . (i
\partial^\perp + gA^\perp) + \gamma^0m \Big ] \psi^+.
\end{eqnarray}
The gauge bosonic part of the Hamiltonian density is given by
\begin{eqnarray}
\theta^{+-}_g  =  - F^{+ \lambda} F^{-}_{\lambda}+ { 1 \over 4} 
g^{+-} (F_{\lambda \sigma})^2 =
{ 1 \over 4} \Big (\partial^+ A^{- }\Big )^2 + 
{ 1 \over 2} F^{ij} F_{ij}
\nonumber \\
 = (\partial^i A^j)^2  
	 + 2e \partial^i A^i \left( \frac{1}{\partial^+}
		\right)  2 (\psi^+)^{\dagger}
		\psi^+ \nonumber \\~~~  + e^2 \left( \frac{1}{\partial^+}
		\right)  2 (\psi^+)^{\dagger}
		 \psi^+  
	  \left( \frac{1}{\partial^+}\right)
		 2 (\psi^+)^{\dagger} 
		\psi^+.   
\end{eqnarray}
The fermionic part of the longitudinal momentum density is given by,
$\theta^{++}_{f} = i \overline{\psi} \gamma^+ \partial^+ \psi$
and the gauge bosonic part of the longitudinal momentum density,
$\theta^{++}_{g} = - F^{+ \lambda } F^-_{\lambda }$.
For a positronium-like bound state, we calculate the matrix element of
$\theta_f^{+-}$ and $ \theta_g^{+-}$. The matrix elements have both diagonal
and off diagonal contribution. The diagonal contribution to the matrix element 
from the fermionic and the gauge bosonic part is given by, 
\begin{eqnarray}
{\Big[\langle P \mid\theta^{+-}\mid P \rangle - {(P^\perp)^2 \over (P^+)^2}
\langle P \mid \theta^{++} \mid P \rangle \Big ]}_{diag}=2\int
dx_1d^2\kappa_1^\perp \psi^*_2\psi_2 \{{(\kappa_1^\perp)^2 \over {x_1}} +
{(\kappa_2^\perp)^2 \over (1- x_1)}\}+\nonumber\\{2\int dx_1d^2\kappa_1^\perp
\int dx_2d^2\kappa_2^\perp \psi_3^*\psi_3 \{{(\kappa_1^\perp)^2 \over {x_1}} +
{(\kappa_2^\perp)\over {x_2}} + {(-\kappa_1^\perp -\kappa_2^\perp)^2 
\over (1-x_1 -x_2)}}\}\nonumber\\
-{8e^2\ov {2(2\pi)^3}} \int dx_1 d^2\k_1^\p \int dy d^2\k^\p
\psi_2^*(x_1,\k_1^\p)\nonumber\\~~~~~~~~~~
 \psi_2(y, \k^\p) {1\ov
{(x-y)^2}}. 
\end{eqnarray}
where $\theta^{+-}=\theta^{+-}_f + \theta^{+-}_g$

The off-diagonal part can be written as,
\be
{ \Big [ \langle P \mid \theta^{+-} \mid P \rangle - {(P^\perp)^2 \over (P^+)^2}
\langle P \mid \theta^{++} \mid P\rangle \Big ]}_{off-diag}
={\cal V}_1 +{\cal V}_2
\e
where,
\be
{\cal V}_1= {2e^2\over {2(2\pi)^3}}   \int dx d^2\kappa^\perp 
\int dy d^2q^\p \Big [ {1\ov E}{1\ov (x-y)} \Big [2V_1
{\mid \psi_2^{\s_1 \s_2} (x,\k^\p) \mid }^2  
\nonumber\\+V_2
(\psi_2^{*\s_1 \s_2} (x,\k^\p)\psi_2^{\sigma_1\sigma_2}(y,q^\p)+ h. c.) \Big
], 
\e

\be
{\cal V}_2= {2e^2\over {2(2\pi)^3}} \int dx d^2\kappa^\perp \int dy
d^2q^\perp
{1\ov E'}{1\ov (y-x)} \Big [
2{V'}_1
{\mid \psi_2^{\s_1 \s_2} (x,\k^\p) \mid }^2 \nonumber\\+{V'}_2
 (\psi_2^{\sigma_1\sigma_2}
(x,\kappa^\perp,1-x,-\kappa^\perp)\psi_2^{\sigma_1\sigma_2}(y,q^\p)+ 
h. c.) \Big ] 
\e
where the expressions for $V_1, V_2, {V'}_1, {V'}_2, E$ and $E'$ are
 given earlier. As before, we have taken the two particle state with both
$\s_1$ ,$\s_2$ up.

Considering only the photon exchange interactions and putting $M^2=M_0^2$ in
the energy denominators as before, one obtains in the non-relativistic limit
for a weak coupling theory, 
\be
 \Big [ \langle P \mid \theta^{+-} \mid P \rangle - &&{(P^\perp)^2 \over (P^+)^2}
\langle P \mid \theta^{++} \mid P\rangle \Big ] = 
2 \int dx d^2\k^\p {\mid
\psi_2 \mid }^2 {{(\k^\p)^2 + m^2}\ov {x(1-x)}} \nonumber\\&&
-2{2e^2\ov {2(2\pi)^3}} \int dyd^2q^\p\psi^*_2( x, \k^\p, 1-x,
-\k^\p) \psi_2(y,q^\p,
1-y,-q^\p) \nonumber\\&&~~~~~~~~~~~~~~~~~~~~~~
\Big [ ({m\ov x})^2 {1\ov { (\k^\p-q^\p)^2 + ({m\ov
x})^2(x-y)^2}} \nonumber\\&&~~~~~~~~~~~~~~~~~~~~~~~~~~~~~+ 
      ({m\ov (1-x)})^2 {1\ov { (\k^\p-q^\p)^2
+ ({m\ov (1-x)})^2(x-y)^2}}\Big ].
\label{d3}
\e
 We can see that 
the right hand side of the above equation, which is nothing but Coulomb interaction,  
has exactly the  same form as 
the interaction part of $\int_0^1 {{ F_L^{\tau=4}(x)}\ov x}$. So, 
the interaction part of $\int_0^1 {{ F_L^{\tau=4}(x)}\ov x}
dx$ can be related to the expectation value of the Coulomb interaction. Since
$\k^\p$ and $q^\p$ are small in the non-relativistic limit, all $(\k^\p)^2$
and 
$(q^\p)^2$ dependence in the numerator of the interaction terms are neglected
compared to the $m^2$ dependent terms. However, the term proportional to
${(\k^\p - q^\p)^2\ov (x-y)^2}$ cannot be neglected because both $x$ and $y$
are almost equal and this term cancels the
instantaneous interaction  in the non-relativistic limit. Both of
these terms originate from $F_{L(g)}$ and one can see that only the gauge
bosonic part of the longitudinal structure function is important for the
Coulomb interaction in the weak coupling limit. The $m^2$ terms in the energy
denominators combine with the other terms to give non-vanishing
contribution. The fermionic part only gives contribution to the kinetic
energy of the fermions. Reminding oneself that we are working in the
light-front gauge and not in the Coulomb gauge, this is a manifestation of
the gauge invariance of the separation of the Hamiltonian density into a
fermionic and a gauge bosonic part.

We introduce  the three vector $\vec{p}$ as, 
$\vec{p}= (\k, \k_z)$ 
where $\k_z$ is defined through a coordinate transformation from $x \in
[0,1]$ to $\k_z \in [ -\infty, \infty]$ by,
$x \equiv {1\ov 2} + {\k_z\ov { 2\sqrt { \k^{\p2} + \k_z^2 +m^2}}}$.
We introduce the bound state wave function $\phi(\vec{p})$ which is
normalized as,
$\int d^3\vec{p}\phi^*(\vec{p})\phi(\vec{p})= 1.$
The bound state equation for the positronium in the weak coupling limit can
be written as \cite{billy},
\be
 \Big [ M^2 - 4( (\vec{p})^2 + m^2)\Big ]\phi(\vec{p}) = - {2e^2 \ov
{2(2\pi)^3}} \int d^3 \vec{p} \phi(\vec{p'}) 
{4m \ov {( \vec{p} - \vec{p'})^2}}. 
\label{d1}
\e

We write  Eq. (\ref{d3})
as,
\be
 \Big [ \langle P \mid \theta^{+-} \mid P \rangle - {(P^\perp)^2 \over (P^+)^2}
\langle P \mid \theta^{++} \mid P\rangle \Big ]  = 2 \int d^3\vec{p} {\mid
\phi(\vec{p})\mid }^2 4 [(\vec{p})^2 + m^2] \nonumber\\~~~~ -{{4e^2}\ov
{2(2\pi)^3}} \int d^3\vec{p}\int d^3\vec{p'} \phi^*(\vec{p}) \phi(\vec{p'})
{4m\ov ( \vec{p} - \vec{p'})^2 }.
\e 
We can now see the more familiar form of the Coulomb interaction.
Multiplying the bound state equation Eq. (\ref{d1}) by $\phi^*(\vec{p})$ and
integrating we get,
\be
M^2 =  \int d^3\vec{p} {\mid
\phi(\vec{p})\mid }^2 4 [(\vec{p})^2 + m^2]  -{{2e^2}\ov
{2(2\pi)^3}} \int d^3\vec{p}\int d^3\vec{p'} \phi^*(\vec{p}) \phi(\vec{p'})
{4m\ov ( \vec{p} - \vec{p'})^2 }.
\label{d4}
\e
Hence from Eq. (\ref{c5}), (\ref{d3}) and (\ref{d4}) it can be seen that the
sum rule \cite{fl} is satisfied in the  weak coupling limit
for a positronium target in light-front QED and it can be written as,
\be
\int_0^1 {{ F_L^{\tau=4}(x)}\ov x} dx  = {2\ov Q^2}\Big [ 
\langle P \mid \theta^{+-} \mid P \rangle - {(P^\perp)^2 \over (P^+)^2}
\langle P \mid \theta^{++} \mid P\rangle \Big ]
= 4{M^2\ov Q^2}.
\e
In the non-relativistic limit for a weak coupling theory,
$M^2 = 4m^2 + 4mB_e$ 
where $B_e$ is the binding energy of positronium. 

From Eq. (\ref{d4}) we obtain,
\be
B_e =  \int d^3\vec{p} {\mid
\phi(\vec{p})\mid }^2 {(\vec{p})^2\ov m} -{{2e^2}\ov
{2(2\pi)^3}} \int d^3\vec{p}\int d^3\vec{p'} \phi^*(\vec{p}) \phi(\vec{p'})
{1\ov ( \vec{p} - \vec{p'})^2}.
\e
The first term in the right hand side is the kinetic energy with ${m\over 2}$ being the reduced
mass of the two body system and the second term is the expectation value of
the Coulomb interaction. So we see that in the weak coupling limit, the sum
rule reduces to a relation connecting the kinetic and potential energies to
the binding energy.
\vskip .2in
{\bf 3. $F_L^{\tau=4}$ for the ground state of positronium}
\vskip .2in
The bound state equation Eq. (\ref{d1}) can be analytically solved for QED,
which is the primary motivation for studying QED. The ground state wave
function of positronium is given by,
\be
\phi_{\nu,s_e,s_{e'}}( \vec{p},s,s') = \phi_{\nu}(\vec{p}) \delta_{s_e,s}
\delta_{s_{e'},s'}
\e
where $s_e$ and $s_{e'}$ label the spin quantum numbers of the electron and
positron respectively and $\nu$ denotes all the other quantum numbers, $\nu
= n,l,m$ correspond with the standard non-relativistic quantum numbers of
hydrogen. 
The spin part factorizes out and the wave function is normalized to $1$.
The wave function is given by,
\be
\phi_{\nu}(\vec{p}) = {4(e_n)^{5\ov 2}\ov ((e_n)^2 +
(\vec{p})^2)^2}Y_{\nu}(\Omega_p)
\e

where
$e_n = {m\alpha\ov {2n}}$ and $Y_\nu(\Omega_p) = Y_{n,l,m}(\Omega)$
are Hyperspherical harmonics. Here $0 \leq \mid m \mid \leq l \leq n-1$.

For 1s state of positronium, we have,
$Y_{1,0,0} = {1\ov {\sqrt {2{\pi}^2}}}$.
In terms of $x$ and $\k^\p$, the 1s state wave function can be written as,
\be
\phi_2(x,\k^\p) = \sqrt{m\ov {\pi}^2} {4(e_1)^{5\ov 2}\ov {\Big [ (e_1)^2 - m^2 +
{1\ov 4} { {(\k^\p)^2 + m^2} \ov {x(1-x)}} \Big ]}^2}
\e
which agrees with \cite{brod} for non-relativistic $x \simeq {1\ov 2}$.

The weak coupling limit contributions to the structure functions $F_2(x)$ and
$F_L^{\tau=4}(x)$ can be directly evaluated using this wave function.
\be
F_2(x)&& = \int d^2\k^\p {\mid \phi_2(x,\k^\p)\mid }^2
\nonumber\\&& = \int d^2\k^\p {Ax^4(1-x)^4 \ov {\Big [ m^2[(1-2x)^2 +
{\alpha}^2x(1-x)] + (\k^\p)^2\Big ]}^4} 
\e
where $A=4.78 \times {10}^{-12} (Mev)^6$.
The integral is convergent and can be evaluated analytically 
introducing a cutoff
$\Lambda$. We get,
\be
F_2(x) = 28.15 \times {10}^{-11} {x^4(1-x)^4\ov {\Big [ (1-2x)^2 +
{\alpha}^2 x(1-x) \Big ]}^3}.
\label{f2}
\e
$F_2(x)$ is very sharply peaked at $x={1\ov 2}$. 

The twist four longitudinal structure function is given by,
\be
{F_L^{\tau=4}\ov x}&& = {4\ov Q^2} \int d^2\k^\p {(\k^\p)^2 \ov {x(1-x)}}
{\mid \phi_2 \mid }^2
\nonumber\\&&
={4\ov Q^2} \int d^2\k^\p {{A(\k^\p)^2 x^3 (1-x)^3 }\ov
{\Big [ m^2 [(1-2x)^2 + {\alpha}^2 x(1-x)] + (\k^\p)^2 \Big ]}^4}.
\e
Here we have considered only the $(\k^\p)^2$ dependent part, the integral of
which is directly connected to the kinetic energy of the electron-positron
pair. The above form is similar to the widely used phenomenological ansatz
for the twist four distribution \cite{ellis} but not exactly the same.  
Evaluating the integral analytically, we get,
\be
{F_L(x)\ov x} = 14.7 \times {10}^{-11} {1\ov Q^2}{x^3(1-x)^3\ov {\Big [
(1-2x)^2 + {\alpha}^2 x(1-x) \Big ]}^2}.
\e 
$Q^2 {F_L(x)\ov x}$ is maximum at $x={1\ov 2}$ and it is also sharply peaked. 
The structure function calculated above (\ref{f2}) falls faster than that
expected from constituent counting rule near $x \rightarrow 1$ \cite{black}. This is
because the non-relativistic wave function behaves differently in the
asymptotic region. It can be seen that it behaves as ${1\over (k^\perp)^4}$
for large $k^\perp$ whereas the relativistic wave function behaves as
${1\over (k^\perp)^2}$ \cite{mu}. The asymptotic behavior of the form factor of
positronium $F(Q^2)$ can be calculated analytically using the above wave function. It
can be shown that $F(Q^2)\sim {1\over Q^4}$ when $Q^2$ is large, as
expected for positronium in the non-relativistic limit \cite{mu}. This
means that it does not have a monopole behavior in this limit and one
cannot expect the constituent counting rule to hold.

To summarize, in this work, we have investigated the twist-four longitudinal structure
function for a positronium-like bound state in light-front QED in the weak
coupling limit. In this limit, we get expressions that look similar to the
familiar non-relativistic expressions, but the entire calculation is fully
relativistic in the leading order in light-front bound state perturbation
theory. We have explicitly verified a sum rule for $F^{\tau=4}_L$ that we
previously proposed.  
 It is worth mentioning here that in the non-perturbative
context, we have investigated before the twist four longitudinal structure function
for a meson in $1+1$ dimension and showed that the sum rule is satisfied
using t'Hooft's equation. In this work, we have shown that in the weak coupling
limit, the sum rule reduces to a relation connecting the kinetic and the
potential energies to the binding energy of positronium. We have also shown
that, in this limit, the fermionic part of  $F^{\tau=4}_L$ contributes only
to the kinetic energy of the fermions and not to the interactions. The twist
four $F_L$ in this limit looks similar to a much used phenomenological
ansatz, however, here we get the result directly from field theory.     
The structure function falls faster than that predicted by constituent
counting rule, as expected.

The twist four part of the longitudinal structure function is important
since it is the leading non-perturbative contribution to $F_L$. The leading
twist contribution to  $F_L$ is perturbative, in contrast to the case of
$F_2$. This analysis for a bound state in weak-coupling light-front QED is
quite interesting since it gives an idea of what goes in such a calculation
in light-front QCD.
Similarity renormalization upto $O(e^2)$ does not produce any
additional interaction in the effective QED Hamiltonian and in the weak
coupling limit, one can work with the canonical Hamiltonian. 
The overall computational framework in QCD is the same
and this analysis in light-front QED allows an analytic understanding of the
problem. 

I would like to thank Prof. A. Harindranath for suggesting this problem and
also for many illuminating discussions.

\end{document}